\documentclass[apj]{emulateapj}
\usepackage{apjfonts}

\newcommand{\msun} {$M_\odot$}
\newcommand{\lp} {LP400$-$22}
\shorttitle{THE RUNAWAY WHITE DWARF LP400$-$22}
\shortauthors{KILIC ET AL.}

\begin{document}

\title{The Runaway White Dwarf LP400$-$22 Has a Companion}

\author{Mukremin Kilic\altaffilmark{1,2}, Warren R. Brown\altaffilmark{1}, Carlos Allende Prieto\altaffilmark{3}, B. Swift\altaffilmark{4}, S. J. Kenyon\altaffilmark{1},\\ J. Liebert\altaffilmark{4}, and M. A. Ag\"{u}eros\altaffilmark{5}}

\altaffiltext{1}{Smithsonian Astrophysical Observatory, 60 Garden Street, Cambridge, MA 02138}
\altaffiltext{2}{Spitzer Fellow; mkilic@cfa.harvard.edu}
\altaffiltext{3}{Mullard Space Science Laboratory, University College London, Holmbury St. Mary, Surrey RH5 6NT, UK}
\altaffiltext{4}{Steward Observatory, University of Arizona, 933 North Cherry Avenue, Tucson, AZ 85721}
\altaffiltext{5}{NSF Astronomy and Astrophysics Postdoctoral Fellow; Columbia University, Department of Astronomy, 550 West 120th Street, New York, NY 10027}

\begin{abstract}

We report the detection of a radial velocity companion to the extremely low mass white dwarf \lp.
The radial velocity of the white dwarf shows variations with a semi-amplitude of 119 km s$^{-1}$ and a 0.98776 day period,
which implies a companion mass of $M\geq0.37$\msun. The optical photometry rules out a main sequence companion. Thus
the invisible companion is another white dwarf or a neutron star.
Using proper motion measurements and the radial velocity of the binary system, we find that it has an unusual Galactic orbit.
\lp\ is moving away from the Galactic center with a velocity of $396 \pm 43$ km s$^{-1}$, which is very difficult to
explain by supernova runaway ejection mechanisms. Dynamical interactions with a massive black hole like that in the Galactic
center can in principle explain its peculiar velocity, if the progenitor was a triple star system comprised of a close binary
and a distant tertiary companion. Until better proper motions become available, we consider
\lp\ to be most likely a halo star with a very unusual orbit.

\end{abstract}


\section{INTRODUCTION}

In recent years, the number of known white dwarfs (WDs) has grown significantly.
Optical spectroscopy shows that most of these objects are hydrogen-atmosphere
WDs with a mass distribution that peaks at $0.6 M_\odot$ \citep{eisenstein06,kepler07}. 
Among the new WDs, there are also a handful of extremely low mass (ELM) WDs with $M \leq 0.3 M_\odot$
\citep{liebert04,eisenstein06,kilic07a}. ELM WDs are rare, comprising $\leq$0.2\% of spectroscopically
confirmed WDs. More importantly, single star evolution cannot produce such low mass WDs in the age of the Galaxy.
Thus, these WDs yield interesting tests of stellar evolution theory.

ELM WDs must undergo significant mass loss during their formation. In one scenario,
they form in close binaries whose evolution includes a phase of mass transfer, during which much of the
WD progenitor's envelope is removed. This prevents a helium flash in the progenitor's core and results in the observed
low-mass, helium-core WD \citep[e.g.,][]{marsh95}. 
Existing observations of ELM WDs do not detect the photometric excess or the
spectroscopic signature expected from main sequence companions.
Hence, the binary companions of known ELM 
WDs are probably either WDs or neutron stars.
This result is consistent with spectroscopic studies of large samples of hydrogen-atmosphere WDs, which conclude that in the majority
of cases low-mass WDs are likely to have degenerate companions \citep[e.g.][]{liebert05,nelemans05}.
ELM WDs are also frequent companions to millisecond pulsars,
although in these radio-selected systems the WDs are frequently too faint for optical spectroscopy to confirm that they
are indeed WDs \citep[cf.\ discussion in][]{vankerkwijk05}. 

In another scenario, low-mass WDs
form from the evolution of single, metal-rich stars. \citet{kilic07c} estimate that the binary fraction for WDs with
$M \sim 0.4 M_\odot$ is $50\%$. They also predict that the binary fraction rises to $100\%$ for WDs with $M < 0.2 M_\odot$,
since such extreme mass loss rates are not expected even for the most metal-rich stars in the Galaxy. Spectroscopic radial
velocity studies of the newly discovered low-mass WDs are therefore essential if we are to discriminate between these mass
loss scenarios by measuring the binary fraction of low-mass WDs and/or to characterize the currently unseen companions.

A radial velocity study of SDSS J091709.55+463821.8 (hereafter SDSS J0917+46), the lowest gravity WD currently known, revealed
velocity variations with an orbital period of 7.6 hr. The companion of SDSS J0917+46 is most likely another WD, although a neutron
star companion is not completely ruled out \citep{kilic07b}.

In this Letter, we describe a radial velocity study of another ELM WD, \lp\ (also known as WD 2234+222 and NLTT 54331).
This star is interesting because of its low mass ($\approx0.17$\msun) and also because of its high tangential velocity
\citep[$> 400$ km s$^{-1}$,][]{kawka06}. \lp\ is the only high-velocity ELM WD currently known. Understanding its
origin is important for understanding the binary star evolution that results in ELM WDs. Our observations are discussed
in \S 2, while an analysis of the spectroscopic data and the discovery of a companion are discussed in \S 3. The nature of the
companion is discussed in \S 4.

\section{OBSERVATIONS}

We used the 6.5m MMT telescope equipped with the Blue Channel Spectrograph to obtain moderate
resolution spectroscopy of \lp\ six times on UT 2008 September 23, four times on September 24, and
three times on December 22.
The spectrograph was operated with the 832 line mm$^{-1}$ grating in second order, providing a wavelength
coverage of 3600 $-$ 4500 \AA. All spectra were obtained with a 1.0$\arcsec$ slit yielding a resolving power of $R=$ 4300.
Exposure times ranged from 5 to 10 minutes and yielded a composite spectrum with a signal-to-noise ratio $S/N > 100$ in
the continuum at 4000 \AA.
All spectra were obtained at the parallactic angle, and comparison lamp exposures were obtained after every exposure.
We checked the stability of the spectrograph by measuring the radial velocity of the Hg emission line at 4358.34\AA\, and
found it to be stable to within 3 km s$^{-1}$.
The spectra were flux-calibrated using blue spectrophotometric standards \citep{massey88}.

Heliocentric radial velocities were measured using the cross-correlation package RVSAO \citep{kurtz98}.
We obtained preliminary velocities by cross-correlating the observations with bright WD templates of known velocity.
However, greater velocity precision comes from cross-correlating \lp\ with itself.
Thus we shifted the individual spectra to rest-frame and summed them together into a high S/N template spectrum.
Our final velocities come from cross-correlating the individual observations with the \lp\ template, and are presented in Table 1.
The errors in velocities are estimated from the cross-correlation peak.
In order to check these error estimates, we added noise to each spectra and performed the cross-correlation
100 times. The errors derived from this analysis are consistent with those returned by the RVSAO package
cross-correlation.

\begin{deluxetable}{cc}
\tablecolumns{2}
\tablewidth{0pt}
\tablecaption{Radial Velocity Measurements for LP400$-$22}
\tablehead{
\colhead{HJD}&
\colhead{Heliocentric Radial Velocity}\\
 & (km s$^{-1}$)
}
\startdata
2454732.63362 & $-$280.35 $\pm$ 5.35 \\
2454732.63772 & $-$296.30 $\pm$ 5.97 \\
2454732.67339 & $-$255.55 $\pm$ 3.63 \\
2454732.73553 & $-$233.37 $\pm$ 3.61 \\
2454732.82065 & $-$161.73 $\pm$ 5.87 \\
2454732.90224 & $-$100.40 $\pm$ 5.93 \\
2454733.60017 & $-$285.16 $\pm$ 5.46 \\
2454733.63906 & $-$270.94 $\pm$ 6.64 \\
2454733.74270 & $-$213.23 $\pm$ 3.81 \\
2454733.90517 & $-$104.57 $\pm$ 5.59 \\
2454822.55410 & $-$269.00 $\pm$ 4.65 \\
2454822.62387 & $-$225.91 $\pm$ 7.36 \\
2454822.67241 & $-$188.40 $\pm$ 5.40
\enddata
\end{deluxetable}

Noise self-correlation is not an issue for our velocities. As a first test, we created a series of co-added templates always
excluding the particular spectrum to be correlated from the co-addition. We found that these velocities are consistent
with those presented in Table 1 within 1 km s$^{-1}$. 
As an additional test, we also used the best-fit WD model spectrum (see \S 3) to
measure radial velocities, and found that the results are consistent within 7 km s$^{-1}$. 
Finally, an independent analysis by one of the authors found radial velocity differences of up to 8 km s$^{-1}$
for individual spectra.
Thus, the systematic errors in our measurements are less than 10 km s$^{-1}$;
the mean velocity difference between the analyses is $0 \pm 5$ km s$^{-1}$. This gives us confidence
that the velocities given in Table 1 are reliable.

\section{LP400$-$22 AND ITS COMPANION}

The radial velocity of \lp\ varies by as much as 196 km s$^{-1}$ between different observations,
revealing the presence of a companion object. We weight each velocity by its associated error and solve for the best-fit
orbit using the code of \citet{kenyon86}. The heliocentric radial velocities are best fit
with a circular orbit and a radial velocity amplitude K = 118.7 $\pm$ 14.1 km s$^{-1}$.
The best-fit orbital period is 0.98776 $\pm$ 0.0001 days with spectroscopic conjunction
at HJD 2454732.81 $\pm$ 0.029. However, several aliases separated by roughly 0.01 day, e.g. 0.9770 $\pm$ 0.0001
and 0.9988 $\pm$ 0.0001 day, are also present.
Figure 1 shows the observed radial velocities and the best fit period for \lp.
Even though we observed \lp\ over 3 nights separated by 90 days, due to its nearly one day orbit, we were only able to cover
half of the orbital phase. 
The long time baseline helps us constrain the orbital period accurately. However, the lack of full orbital
coverage causes the relatively large error (12\%) in the velocity semi-amplitude measurement.

\begin{figure}
\includegraphics[angle=0,scale=.45]{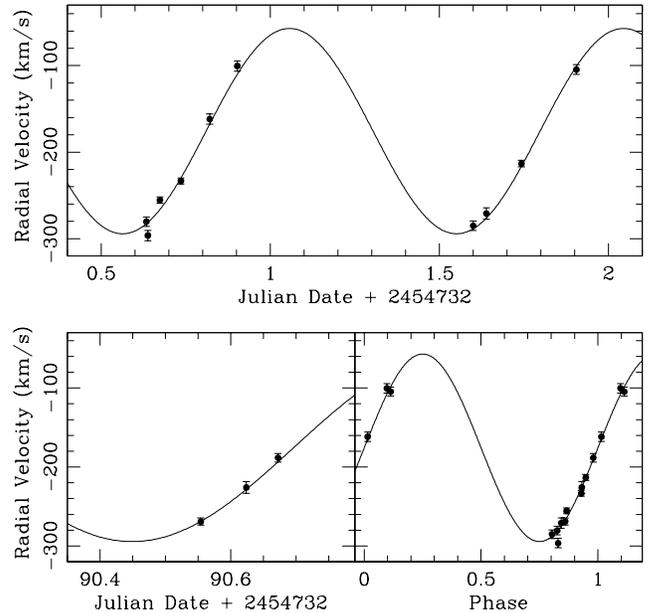}
\caption{The radial velocities of the white dwarf \lp\ (black dots) observed in 2008 September (top panel) and 2008 December (bottom left panel).
The bottom right panel shows all of these data points phased with the best-fit period. The solid line represents the best-fit model for a circular orbit with a radial
velocity amplitude of 118.7 km s$^{-1}$ and a period of 0.98776 days.}
\end{figure}

The discovery spectra of \lp\ from the APO 3.5m telescope were kindly made available to us by A. Kawka. These data consist of
two exposures obtained in 2001 and have different resolution and wavelength coverage from our observations \citep[see][]{kawka06}. 
Therefore systematic differences in measuring the radial velocities are inevitable. Cross-correlating these two spectra with our
template spectrum, we measure velocities of $-177.1 \pm 14.3$ km s$^{-1}$ and $-32.2 \pm 8.4$ km s$^{-1}$. Including these measurements
in our orbital fits changes the orbital period slightly to 1.01 day, but with a significantly larger $\chi^2$. The velocities
from these spectra are also consistent with the range of velocities expected from our best-fit orbital solution. We note that
the choice of periods mentioned above makes negligible changes to the radial velocity semi-amplitude.

We perform model fits to each individual spectrum and also to the composite spectrum using synthetic WD spectra kindly
provided by D. Koester. We use the 13 individual spectra to obtain a robust estimate of the errors in our analysis.
Figure 2 shows the composite spectrum and our fits using the entire spectrum and also using only the Balmer lines.
A best-fit solution of $T_{\rm eff}=11440 \pm 70$ K and $\log$ g = 6.35 $\pm 0.01$ results from the observed composite spectrum. 
Slight differences between the continuum level of the observations and that of the best-fit model spectrum
redward of 4000 \AA\ show that the flux calibration
was not perfect. If we normalize (continuum-correct) the
composite spectrum and fit just the Balmer lines, then we obtain $T_{\rm eff} =11290 \pm 50$ K and $\log$ g = 6.30 $\pm 0.02$. 
Our results are consistent with each other, and also with Kawka et al.'s (2006) estimates of
$T_{\rm eff} = 11080 \pm 140$ K and $\log$ g = 6.32 $\pm 0.08$.
We adopt our best fit solution of $T_{\rm eff} = 11290 \pm 50$ K and $\log$ g = 6.30 $\pm 0.02$ for the remainder of the paper.
We confirm that \lp\ is an ELM WD.

\begin{figure}
\includegraphics[angle=-90,scale=.35]{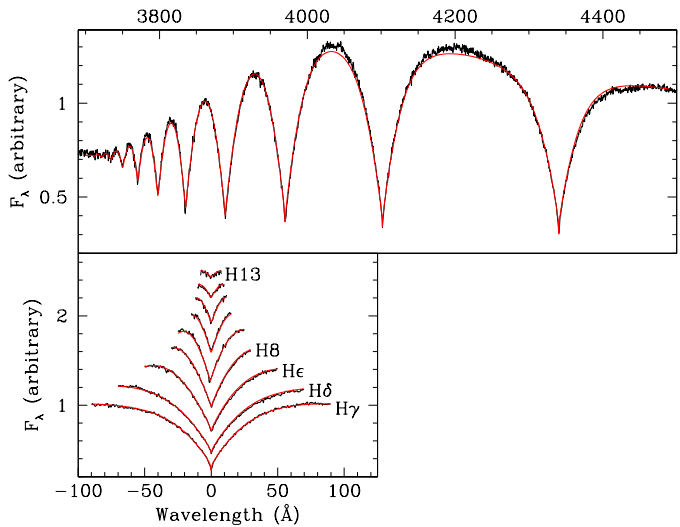}
\caption{Spectral fits (solid lines) to the observed composite spectrum of \lp\ (jagged lines, top panel) and to
the flux-normalized line profiles (bottom panel).}
\end{figure}

Comparing our temperature and surface gravity measurements to \citet{althaus01} models shows that \lp\ has $M\approx0.17$\msun.
The effective temperature and surface gravity estimates for \lp\ are slightly different than the predicted values for a
0.17\msun\ WD \citep[see Fig. 9 in][]{kilic07a}. \citet{kawka06} used their best-fit model spectra and the mass-radius relations of \citet{althaus01} and
\citet{serenelli01} to estimate an absolute magnitude of $M_V=9.1 \pm 0.2$ mag, a distance of 430 $\pm$ 45 pc, and a WD 
cooling age of 500 Myr. We adopt these values for our analysis as well.
Using the orbital period and the semi-amplitude of the radial velocity variations, we estimate a mass function for \lp\ of
0.171 $\pm$ 0.043. Using $M=0.17$\msun\ for the WD, we can set a lower limit on the mass of the companion by assuming
an edge-on orbit (sin $i=$ 1), for which the companion would be an 0.37\msun\ object at an orbital separation of 3.4$R_{\sun}$.
Therefore, the companion mass is $M\geq0.37$\msun.

\section{THE NATURE OF THE COMPANION}

\subsection{A Low Mass Star}

We combine the spectra near maximum blue-shifted radial velocity and near minimum radial velocity into two composite spectra. If there
is a contribution from a companion object, it may be visible as an asymmetry in the line profiles.
We do not see any obvious asymmetries in the line profiles and conclude that our optical spectroscopy does not reveal any spectral
features from a companion object.

\lp\ has $M_V\approx M_I\approx 9.1$ mag \citep{kawka06}.
A low mass star companion with $M\geq0.37$\msun\ would have $M_I<8.8$ mag \citep{kroupa97}, brighter
than the low mass WD and detectable in the $I$-band. Hence, a main sequence star companion is ruled out.

\subsection{Another White Dwarf}

Using the mean inclination angle for a random stellar sample, $i=60^{\circ}$, we estimate that the companion mass is
probably $\sim$0.48\msun.
\citet{kilic07b} studied possible formation scenarios for the SDSS J0917+46 binary system involving two common envelope phases.
Using the same formalism used in that study \citep[$\gamma$-algorithm equating the angular momentum balance,][]{nelemans05},
we search for possible progenitor masses and binary separations to form \lp\ and a 0.48\msun\ WD companion with an orbital period of 0.98776 day.

The Galactic orbit for \lp\ is most compatible with a halo object (see \S 5). Therefore the main sequence age of the
progenitor star is $\sim$10 Gyr; the progenitor was a $\sim$1\msun\ main sequence star.
We assume that the mass of the WD is the same as the mass of the core of the giant at the onset of the mass transfer.
We assume a giant mass of 0.8-1.2\msun, a core mass of 0.17\msun, and possible WD companion masses of $0.37-1.39$\msun.
We estimate that a common envelope phase involving a 0.9\msun\ star and
a 0.48\msun\ companion at an orbital separation of $6R_{\odot}$ and orbital period of 1.45 days can create the \lp\ binary
observed today. 

The same algorithm can be used to re-create the first common envelope phase. However, we do not find any possible solutions involving
$M<2.3$\msun\ stars\footnote{This limit is set by the fact that more massive stars do not form degenerate helium cores
and that a common envelope phase with a more massive giant would end end up in a merger and not in a binary system \citep{nelemans00}.}
if we use $\gamma=1.5$, where $\gamma$ is the rate of angular momentum loss as defined by \citet{paczynski67}. 
\citet{nelemans05} found that $\gamma=1.5$ can explain most of the systems that they studied, however the
first phase of mass transfer for individual systems could be explained by algorithms with $\gamma \approx0.6-3$ (see their Fig. 1).
Assuming $\gamma=2$ for the first common envelope phase, the \lp\ system
can be explained as the descendant of a 0.9\msun\ star and a $1.8-2.0$\msun\ star with an orbital
separation of 1.9 AU. Therefore, a likely evolutionary scenario for a WD + WD binary involving \lp\ 
is: 2.0\msun\ giant + 0.9\msun\ star at 1.9 AU $\longrightarrow$ 0.48\msun\ WD + 0.9\msun\ star at $6R_{\odot}$ $\longrightarrow$ 0.48\msun\ WD
+ 0.9\msun\ giant at $6R_{\odot}$ $\longrightarrow$ 0.48\msun\ WD + 0.17\msun\ WD at $3.6R_{\odot}$.

The main sequence lifetime of a 2\msun\ star is less than 1 Gyr, and a 0.48\msun\ halo WD created
from such a system is $\sim$10 Gyr old. According to the \citet{bergeron95} models, a 0.5\msun\ CO-core
WD cools down to 3250 K in 10 Gyr and it has $M_I\sim15.4$ mag. 
This companion is several orders of magnitude fainter than the 0.17\msun\ WD observed today, and therefore
the lack of evidence of a companion in the optical photometry is consistent with this formation scenario.

\subsection{A Neutron Star}

If the orbital inclination angle of the \lp\ binary system is less than 32$^{\circ}$, the companion mass is $\geq$1.4\msun, consistent with
a neutron star. Given the observational connection between low-mass WDs
and millisecond pulsars (MSPs), there have been several attempts at detecting MSP companions to the newly discovered low-mass WDs.
\citet{agueros09} conducted a search for pulsar companions to $15$ low-mass WDs spectroscopically identified in the SDSS at $820$ MHz
with the NRAO Green Bank Telescope. However, no convincing pulsar signal was detected in their data, and they conclude that the probability that
the companion to a given low-mass WD is a MSP is $< 10^{+4}_{-2}\%$. In addition, the probability of observing a binary system at an
angle less than 32$^{\circ}$ is only 15\%. We require radio and X-ray observations of \lp\ to put constraints on a possible
pulsar companion (Ag\"{u}eros et al., in preparation).

\section{DISCUSSION}

Our radial velocity measurements of \lp\ show that it is in a binary system with an orbital period of 0.98776 day.
Short period binaries may merge within a Hubble time by losing angular momentum through gravitational radiation. However, the
merger time for the \lp\ binary is longer than 230 Gyr for either a WD or a neutron star companion.

At a Galactic latitude of $-30.6^{\circ}$, \lp\ is $\approx200$ pc below the plane. The systemic radial velocity
of the binary system is $-175.7 \pm 11.4$ km s$^{-1}$, and the proper motion is
\citep[$\mu_{\alpha} cos \delta, \mu_{\delta}) = (198, 53$ mas yr$^{-1}$;][]{lepine05}.
The velocity components\footnote{The velocity components $U, V,$ and $W$ are directed to the Galactic center, rotation direction,
and north Galactic pole, respectively.} with respect to the local standard of rest as defined by \citet{hogg05} are
$U=-396 \pm 43, V=-195 \pm 15$, and $W=-27 \pm 19$ km s$^{-1}$. Clearly, \lp\ is not a disk star.
In the Galactic rest frame its total velocity is 398 $\pm$ 50 km s$^{-1}$.
This is slightly lower than the canonical escape velocity of 500$-$550 km s$^{-1}$
in the solar neighborhood \citep{carney88,smith07}.

For comparison, halo stars within 1 kpc of the Galactic plane have typical velocities of
$U=-17 \pm 141, V=-187 \pm 106$, and $W=-5 \pm 94$ km s$^{-1}$ \citep{chiba00}. 
Thus, \lp's $U$ velocity makes it an outlier among halo stars. 
There is only one star in the \citet{chiba00} study,$-29$ 201W1, that has $UVW$ velocities consistent with those of
\lp\ within 2$\sigma$. This star demonstrates that
stars with similar kinematics to \lp\ do exist, but comprise only 0.1\% of nearby metal poor stars
in the \citet{chiba00} sample. 
We now explore whether \lp\ is more plausibly explained as a disk runaway or a Galactic center ejection.

Runaway disk stars are explained by velocity kicks from 3- or 4-body dynamical interactions or from binary companions that
explode as supernovae. Depending on the orbital separation
and binary mass fraction, a supernova explosion may or may not disrupt the binary system. If the companion to \lp\ is
a neutron star, it could be responsible for the observed high velocity of the system. However, detailed binary population synthesis
calculations of runaway stars by \citet{portegieszwart00} show that
less than 1\% of runaways receive velocity kicks in excess of 200 km s$^{-1}$. 
Known low mass WD + MSP systems can be used to test the supernova kick scenario. 
We estimate that the PSR J1012+5307 and PSR J1911$-$5958A binary systems
have total velocities $\leq90$ km s$^{-1}$, based on
the radial and tangential velocity estimates by \citet{lorimer95}, \citet{bassa06}, and \citet{corongiu06}.
This comparison shows that \lp\ is unique in its large space velocity, and 
the supernova kick scenario is unlikely to explain it.

\citet{justham09} recently suggested that single runaway WDs may form from SNe Ia in short period ($\sim$1 hr) binary systems.
Since we now know that \lp\ is a binary
with an almost one day period, the SNe Ia mechanism is ruled out for this system.

Figure 3 plots the Galactic orbit of \lp\ for the past 1 Gyr, in a static
disk-halo-bulge potential \citep{kenyon08}.
Given its relatively low $W$ velocity, \lp\ stays within 10 kpc of the
Galactic plane. The large $U$ velocity causes it to move mostly in the radial direction, and
its closest approach to the Galactic center occurs at a distance of $R\approx$ 270 pc.
Given the uncertainties in the measured parameters, we estimate that the last pericenter passage occurred around 16 Myr ago
with a 0.1\% chance that \lp\ passed within 10 pc of the Galactic center.
Recent discoveries of unbound hypervelocity stars in the Galaxy suggest that the extreme velocities of these stars
come from dynamical interactions with the massive black hole in the Galactic center \citep{brown09}. 
However, massive black hole ejection mechanisms \citep{hills88,yu03} are much more likely to eject single stars than binaries
\citep{lu07,perets09}, as most of the binaries are not expected to survive close to the massive black hole in the Galactic center.
Tidal disruption of a hierarchical triple star system by a central massive black hole could, in principle, lead to the ejection
of a close binary. \citet{lu07} find that the typical ejection speed of such a binary would be 400 km s$^{-1}$. 
However, all things considered, \lp\ is most likely a halo binary star system with an unusual orbit.

\begin{figure}
\includegraphics[angle=0,scale=.45]{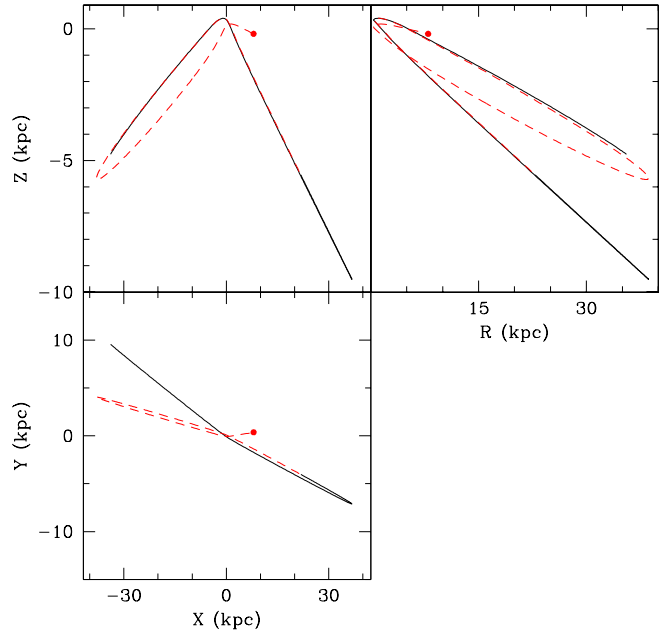}
\caption{The Galactic orbit of \lp\ for the past 1 Gyr. The orbital motion since \lp\ became a WD (500 Myr ago) is shown as
a dashed line. The current position of the WD is marked with a dot.}
\end{figure}

\section{CONCLUSIONS}

The runaway WD \lp\ has a radial velocity companion. The optical photometry
and the orbital parameters of the system rule out a low mass main sequence star companion.
Although a neutron star companion cannot be ruled out, the most likely companion is another WD.
A supernova kick or a Galactic center ejection is unlikely to explain the runaway nature of \lp. 
We suggest that \lp\ may belong to a small sample of halo WDs with unusual orbits.

Excluding the two millisecond pulsar systems, \lp\ is only the second ELM
WD studied for optical radial velocity variations. Both \lp\ and SDSS J0917+46
show radial velocity variations due to compact companions, supporting the binary formation scenario for ELM WDs.
A radial velocity follow-up survey of the other
ELM WDs found in the SDSS is currently underway at the MMT.

\acknowledgements
Support for this work was provided by NASA through the Spitzer Space Telescope Fellowship Program, under an award from Caltech.
M. A. Ag\"{u}eros is supported by an NSF Astronomy and
Astrophysics Postdoctoral Fellowship under award AST 06-02099.

{\it Facilities:} \facility{MMT (Blue Channel Spectrograph)}

\end{document}